# Observation of strange metal in hole-doped valley-spin insulator

Tuan Dung Nguyen[1,2†], Baithi Mallesh[1,2†], Seon Je Kim[2†], Houcine Bouzid[1], Byeongwook Cho[1,2], Xuan Phu Le[1,2], Tien Dat Ngo[3], Won Jong Yoo[3], Young-Min Kim[1,2*], Dinh Loc Duong[1,2*], Young Hee Lee[1,2,4*]

[1]Center for Integrated Nanostructure Physics (CINAP), Institute for Basic Science (IBS), Suwon 16419, Republic of Korea

[2]Department of Energy Science, Sungkyunkwan University, Suwon 16419, Republic of Korea

[3]SKKU Advanced Institute of Nano-Technology (SAINT), Sungkyunkwan University, Suwon 16419, Republic of Korea

[4]Department of Physics, Sungkyunkwan University, Suwon 16419, Republic of Korea

[†]These authors contributed equally to this work

[*] Correspondence to: youngmk@skku.edu, ddloc@skku.edu, leeyoung@skku.edu

**Abstract**

Temperature-linear resistance at low temperatures in strange metals is an exotic characteristic of strong correlation systems, as observed in high-$T_C$ superconducting cuprates, heavy fermions, Fe-based superconductors, ruthenates, and twisted bilayer graphene. Here, we introduce a hole-doped valley-spin insulator, V-doped $WSe_2$, with hole pockets in the valence band. The strange metal characteristic was observed in $V_xW_{1-x}Se_2$ at a critical carrier concentration of 9.5 x $10^{20}$ $cm^{-3}$ from 150 K to 1.8 K. The unsaturated magnetoresistance is almost linearly proportional to the magnetic field. Using the ansatz $R(H,T) - R(0,0) \propto \sqrt{(\alpha k_B T)^2 + (\gamma \mu_B B)^2}$, the γ/α ratio is estimated approximately to 4, distinct from that for the quasiparticles of LSCO, $BaFe_2(As_{1-x}P_x)_2$ (γ/α=1) and bosons of YBCO (γ/α=2). Our observation opens up the possible routes that induce strong correlation and superconductivity in two-dimensional materials with the strong spin-orbit coupling.

Resistance in metals is mainly caused by electron-phonon, electron-electron, and electron-impurity scatterings [1–3]. The scattering rates in the two former mechanisms rely on temperature. The resistance-temperature relationship follows the equation $\rho = \rho_0 + AT^2 + BT^5$ in the low-temperature regime (e.g., below the Debye temperature) for most metals [1–4]. A temperature exponent of 2 is dominant at extremely low temperatures, which is characterized by the electron-electron scattering in Fermi liquids, whereas an exponent of 5 is represented by electron-phonon scatterings. As opposed to the case of normal metals, the T-linear resistance of strange metals is poorly understood [5,6].

The first observation of a strange metal with high-$T_C$ superconducting cuprates brings attention to this exotic phase [7–14]. Since then, strange metal characteristics have been explored in other superconductors such as heavy fermions β-$YbAl_{1-x}Fe_xB_4$ [15], iron-based $BaFe_2(As_{1-x}P_x)_2$ [16,17], ruthenates [18,19], and more recently, twisted bilayer graphene [20–22]. In some cases, however, the strange metal does not show superconductivity under an external force (e.g., pressure or magnetic field) [18,23]. Interestingly, a recent experiment revealed that the strange metal can be observed by bosonic charges (e.g., copper pairs) [12]. Herein, we report for the first time strange metal characteristics in a van der Waals layered material, $WSe_2$ with vanadium as a hole dopant.

We revisited the band structure of normal, spin-$\frac{1}{2}$ Mott-Hubbard, and valley-spin insulators (Fig. 1). In the normal insulator [Fig. 1(a)], the total number of electrons in the atoms of a primitive cell are even numbered (e.g. Si), 2N. If there is no overlap between the fully filled and empty bands, the solid is an insulator. The spin states of the electrons in the primitive cell are indistinguishable. There is no difference between the spin-up and spin-down bands in real and reciprocal spaces. In contrast, the primitive cell of the Mott-Hubbard insulator has one electron per atom. Since the primitive cell has two atoms in Mott-Hubbard insulator, the total number of electrons is even [Fig.

1(b)]. Furthermore, the strong Coulombic repulsion between the atoms (considered as U potential) induces strong antiferromagnetic coupling, splitting a single 2N-state band into two N-state bands for each spin [Fig. 1(b)]. Hence, an insulating band is formed with identical spin-up and spin-down bands in the reciprocal space, but the spins located in different atoms are identical in the real space. This type of band represents a strong correlation system with large Coulombic interactions (U).

Strong spin-orbit coupling (λ) with broken inversion symmetry, for example, in transition metal dichalcogenides, provokes two distinct spin-up and spin-down bands in the reciprocal space, although the number of electrons in the atoms is even. Consequently, four N-state bands with distinct spin states are formed [Fig. 1(c)], called a valley-spin insulator, which is similar to the density of states (DOS) in a spin-$\frac{1}{2}$ Mott-Hubbard insulator. The spin states of the valley-spin insulator are distinguished in momentum space, but those of a Mott-Hubbard insulator are located separately in real space. The similarity of the distinct spin-up and spin-down DOSs between Mott-Hubbard and valley-spin insulators motivated us to investigate the doped valley-spin insulator for verifying the possibilities of the cooper-pair formation. We note that Ising pairing superconductivity, where the cooper pair is induced by intervalley spin-coupling between K and K', was observed in $NbSe_2$ [24]. Since doped Mott-Hubbard insulators exhibit strange metal behavior [9–11,25–27], it will be interesting to investigate such strange metal behavior and its possible superconductivity in degenerately doped transition metal dichalcogenides, for example, p-doped $WSe_2$ with a V-dopant, as a prototype material. The V atom may shift the Fermi level deep down into the edge of the valence band while triggering the impurity band to cross the valence band, inducing interactions between K and K' valleys.

Figure 2(a) shows a ball-and-stick schematic representation of pure and V-doped $WSe_2$ lattice. V atoms are substituted into W atoms. The substituted V atoms are confirmed experimentally [28–31]. The valley-spin pockets are presented in the schematic representation [Fig. 2(b)], where the band structure is calculated by the density functional theory [Fig. 2(c)]. The Fermi level is located deeply inside the valence band at a V doping concentration of 12.5%, representing degenerate states. This type of band is similar to that in $Sr_xLa_{1-x}CuO_4$ [32,33].

To investigate the properties of the suggested doped valley-spin insulator, heavily V-doped $WSe_2$ single crystals were synthesized with various V-doping concentrations by a chemical vapor transport process (CVT), as described in the Methods section. Figure 3(a) shows the XRD pattern of the crystalline 25% V-doped $WSe_2$, revealing sharp peaks at 13.57°, 41.77°, and 56.90° corresponding to the (002), (006), and (008) crystalline orientations, which are well-matched with the 2H-$WSe_2$ reference spectra (ICSD 01-071-0600) . The H phase of the host $WSe_2$ is maintained even at very high V-doping concentrations, consistent with previous reports on heavily monolayer V-doped $WSe_2$ grown by chemical vapor deposition [28]. The appearance of the additional peaks at 31.14° and 37.63° is also consistent with the (100) and (103) peaks of 2H-$WSe_2$. The XRD patterns for other V concentrations are included in Supplemental Material Fig. S1 [34]. The presence of the V atom as a substitutional dopant has been identified in high-resolution TEM images [28,35] and further confirmed by the additional Raman peak near 200 $cm^{-1}$ at 10% V, which becomes more pronounced at 25% V concentration in Fig. 3(b) [36]. The V atoms are uniformly distributed over the sample with 25% V concentration, as described in the intensity mapping of the V peak near 200 $cm^{-1}$.

The H-phase structure is persistent at a V concentration as high as 25% from the W-4*f*, Se-3*d*, and V-2*p* core levels in the X-ray photoemission spectra (XPS) [Fig. 3(c)]. The doublets of W-4*f* (32.65 and 34.84 eV) and Se-3*d* (54.92 and 55.75 eV) core levels of the pristine 2H $WSe_2$ are red-shifted, rational to the V-doping concentration (Fig. S3 [34]). Since the binding energy in the XPS spectrum is the energy difference between the Fermi level to the core level, the redshift of W-4*f* and Se-3*d* indicates that the Fermi level is shifted toward the valence band of $WSe_2$ [37,38], i.e., p-type doping. This observation is also consistent with the DFT band structure. The V-2*p* core-level presents two pairs of V-2$p_{3/2}$ and V-2$p_{1/2}$ peaks centered at (513.7, 521.4 eV) and (516.4, 524.9 eV) in the spectrum of V-doped $WSe_2$, which corresponds to $V^{4+}$ and $V^{5+}$ oxidation states, respectively. The intensity of the V-2*p* core-levels increases with the V-doping concentration (Fig. S3 [34]). The observed W-*4f* and Se-*3p* peaks are located far from that of the 1T'-$WSe_2$ phase [39], confirming the H structure of the doped samples. The H structure of multilayer $WSe_2$ sample doped by 25% V atoms was confirmed by scanning transmission electron microscopy (STEM) [Fig. 3(d)]. The diffraction pattern revealed typical hexagonal spots confirming the H-phase (inset). Consistently, atomic resolution annular dark-field STEM image of the sample clearly shows the lattice structure of H-phase (middle panel). STEM-based energy dispersive X-ray spectroscopy analysis of V $K_\alpha$, W $M_\alpha$, and Se $L_\alpha$ edges in the sample further corroborates the presence of V atoms in the lattice (right panel).

Then, we investigated the electrical properties of heavily V-doped $WSe_2$ samples. Since the V concentration could be different from the expected nominal values, we prepared a Hall bar device and determined the carrier concentration rather than the nominal concentration. Figure 4(a) shows the variation of resistance with temperature. The R-T curves are fitted with $\rho = \rho_0 + AT^\delta$ at low temperatures below 150 K. All the samples revealed a nonquadratic resistance with $\delta$ smaller than 2 within the measured ranges of carrier density (0.7–3 x $10^{21}$ $cm^{-3}$) when the temperature reached 1.8 K, revealing a non-Fermi liquid-like nature. In particular, the sample with a carrier concentration of 9.46 x $10^{20}$ $cm^{-3}$ manifests a linear R-T curve from 1.8 K to 150 K, i.e., it exhibits strange metal behavior. Such a strange metal characteristic in a hole-doped valley-spin insulator without superconductivity is distinct from the mixture of superconductivity and strange metal in other systems [10,13,15,16,21], enriching the material family (Table 1).

The $\delta$ value of the power is plotted with different temperatures and carrier densities [Fig. 4(b)]. The color map indicating the variation of $\delta$, manifests two non-Fermi liquid regimes, separated by a critical region with $\delta$=1 at an approximate carrier density of 9.46 x $10^{20}$ $cm^{-3}$ at low temperature. This phase diagram with a possible quantum critical point is similar to the other critical transitions observed by controlling the doping concentration, pressure, and magnetic field [18,23,40,41]. As the magnetic field along c-axis increases [Fig. 4(c)], the strange metal behavior at a carrier density of 9.46 x $10^{20}$ $cm^{-3}$ at a low temperature is suppressed with increasing resistance. This is due to the dominant effect of the magnetic field on the resistance at low temperature, which induces positive magnetoresistance [16,19,42]. Similar behavior was exhibited by several superconducting materials such as cuprates and iron pnictide [16,43].

We then investigated the effect of magnetic field on the samples with T-linear resistance [Fig. S4 [34], Figs. 4(d) and 4(e)]. The magnetoresistance reveals a linear relation between resistance and magnetic field up to the measurement limits of 8T, similar to other strange metal systems [12,42–44]. The universal T- and B-linear resistance yields the ansatz:

$$R(H,T) - R(0,0) \propto \sqrt{(\alpha k_B T)^2+(\gamma \mu_B B)^2} \quad [16]$$

where α and γ are constant numbers, which can be estimated by the slope extracted from R ($k_B T$) at B=0 [Fig. 4(d)] and R ($\mu_B \mu_0 H$) at T=1.8K [Fig. 4(e)]. Interestingly, the γ/α ratio in our samples is approximately 4, which is different from that of the fermions (γ/α = 1) in LSCO, $BaFe_2(As_{1-x}P_x)_2$ and bosons (γ/α = 2) in YBCO. The distinct γ/α in the strange metal of $V_xW_{1-x}Se_2$ is attributed to another coupling mechanism with strong spin-orbit coupling in $WSe_2$, which plays an important role in their electrical transport properties.

**Summary**

We have observed the strange metal behavior in V-doped valley-spin insulator $WSe_2$. Although superconductivity was not observed in the strange metal V-doped $WSe_2$, as opposed to the case of doped cuprates and iron-based superconductors, our observation opens up a new path for the search of novel superconducting low-dimensional materials with strong SOC. It will be interesting to investigate similar dopants such as Nb and Ta for $WSe_2$. Other host materials such as $WS_2$, $MoS_2$, and $MoSe_2$ can be good candidates. Furthermore, to enhance the degrees of freedom that control the properties of hole-doped TMDs, for example, the Fermi level, a co-doping strategy can be adopted by adding an n-type dopant such as Re.

**Acknowledgements** This work was supported by the Institute for Basic Science (IBS-R011-D1), Republic of Korea; Advanced Facility Center for Quantum Technology; Advanced Facility Center for Quantum Technology. Y.M.K. acknowledges the financial support by the Basic Science Research Program through the National Research Foundation of Korea (NRF) funded by the Ministry of Science, ICT & Future Planning (2020R1A4A4078780). The use of the TEM instrument was supported by Advanced Facility Center for Quantum Technology in SKKU.

## Figures

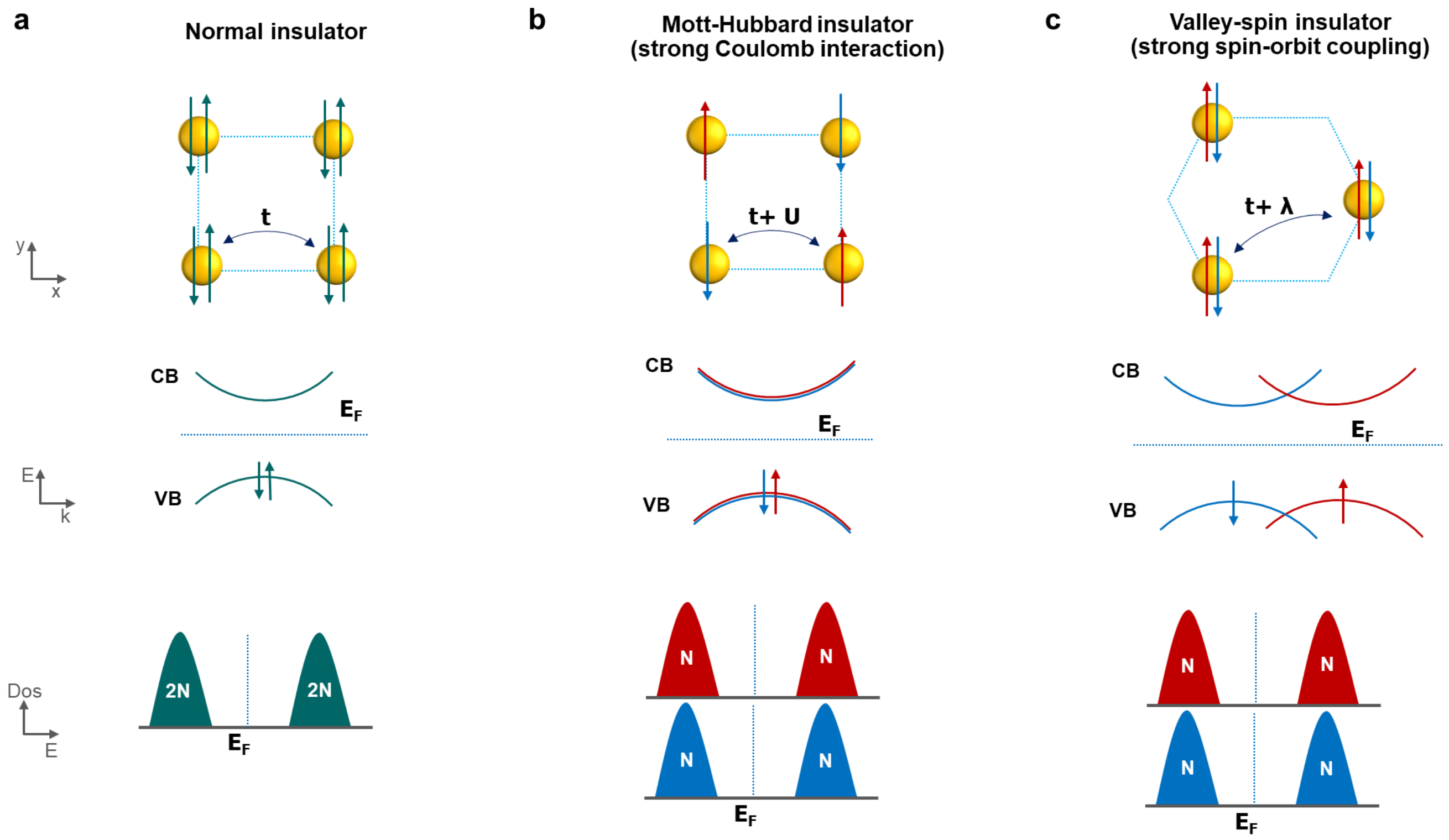

FIG. 1. Schematic band structures of different types of insulators in real and reciprocal space. (a) Normal insulator with two-electron atom in primitive cell. (b) Spin-1/2 Mott-Hubbard insulator with two single-spin atoms in primitive cell, which forms an antiferromagnetic order. Spin density is locally disguisable although spin-up and spin-down bands in reciprocal space are identical. (c) Valley-spin insulator with two-electron atom in primitive cell. No spin density appears in real space. Spin-up and spin-down bands, however, are distinguishable in reciprocal space due to strong spin-orbit coupling. Both Mott-Hubbard and valley-spin insulators induce distinguishable spin-up and spin-down density of states.

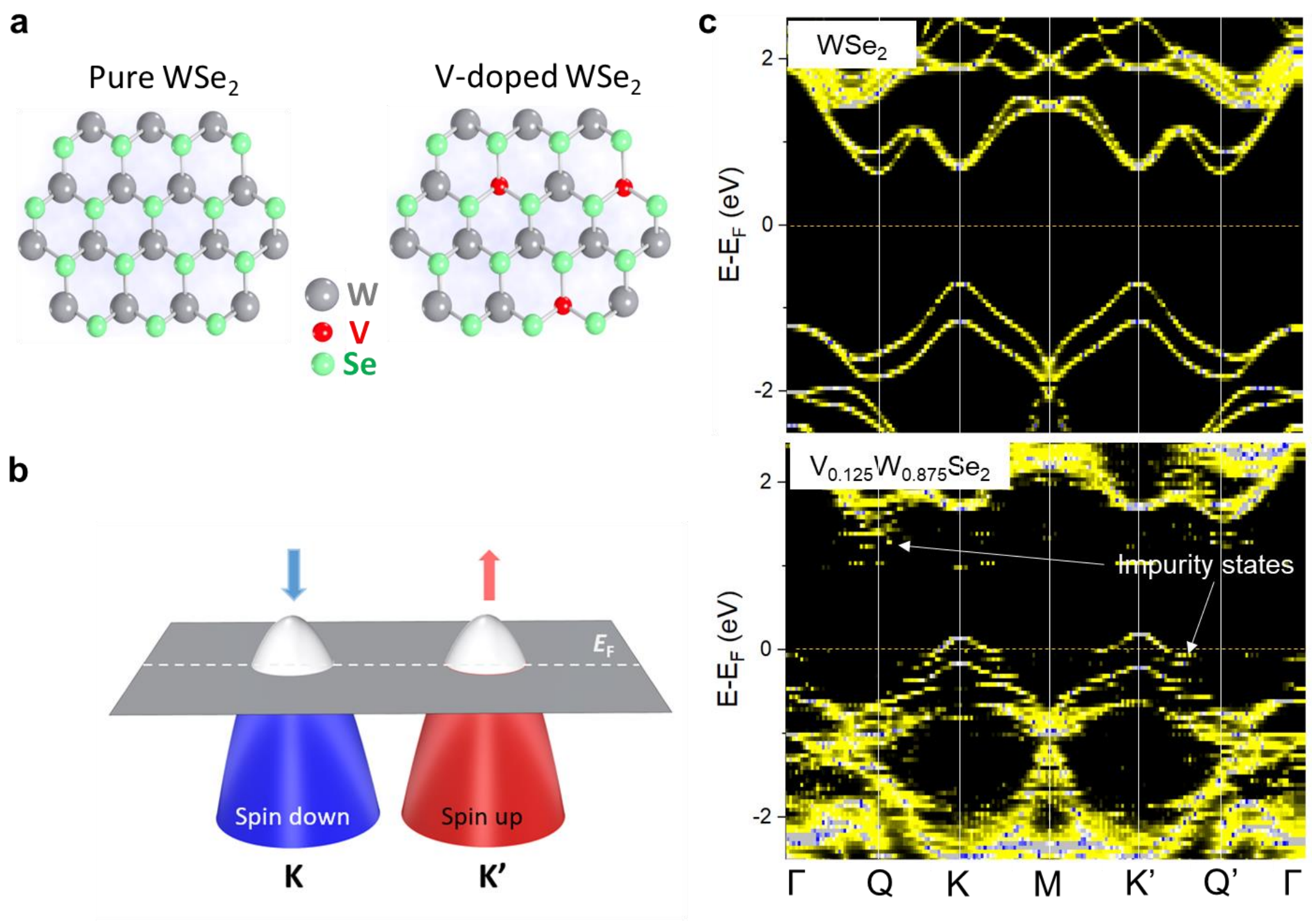

FIG. 2. Atomic and band structures of V-doped $WSe_2$. (a) Schematic illustration of the crystal structure of pure and V-doped $WSe_2$ containing V-substituted atoms in W sites. (b) Schematic of hole-doped valley-spin insulator with distinct spin states in momentum space. (c) Projection density of states in reciprocal space of the pristine $WSe_2$ and V-doped $WSe_2$. The Fermi level (orange line) is shifted to the valence band, revealing the hole pocket at k-points of K and K'. Vanadium concentration is 12.5%.

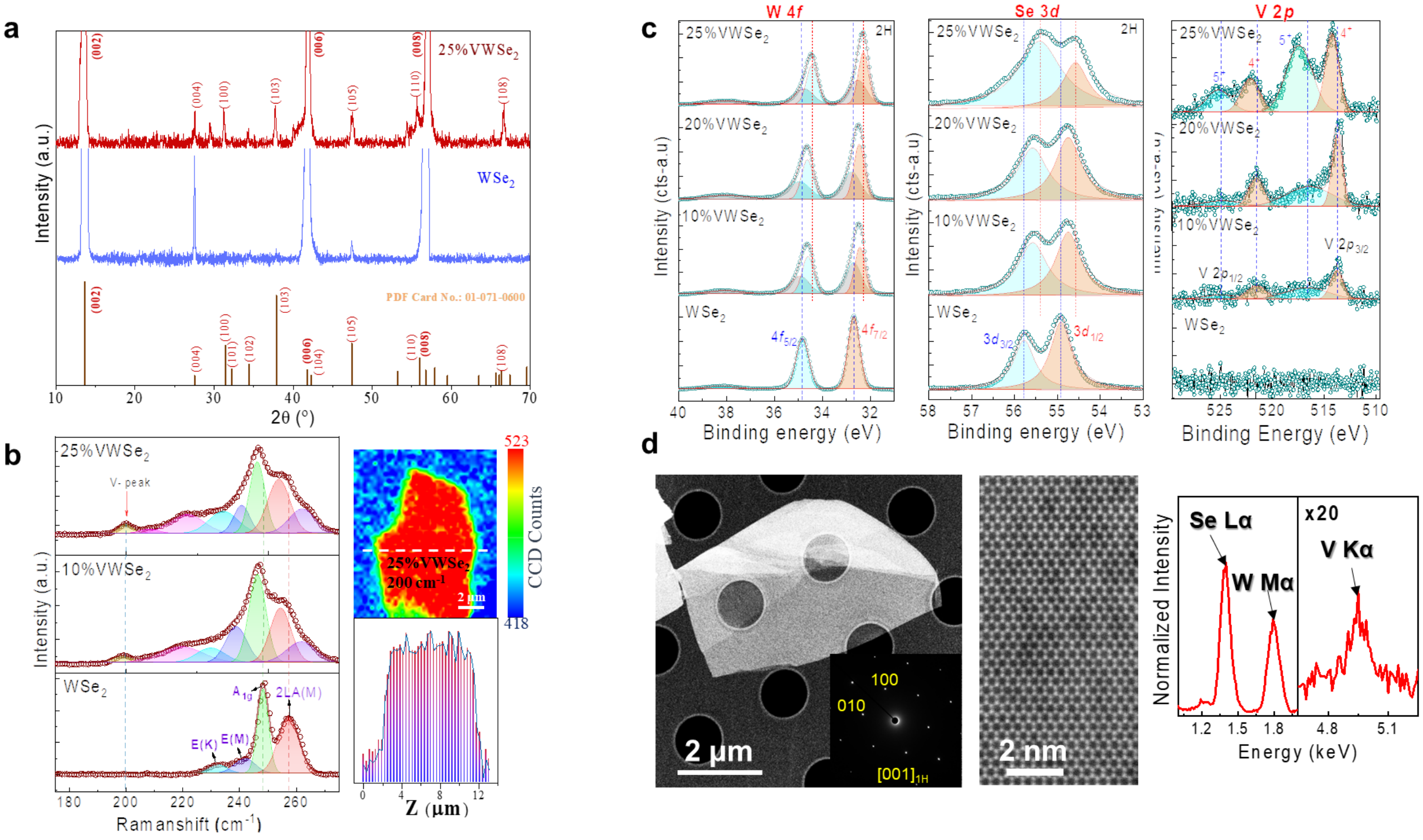

FIG. 3. Crystal structure characterization of V-doped $WSe_2$. (a) X-ray diffraction patterns of pure and 25% V-doped $WSe_2$ with the reference card of 2H-$WSe_2$. (b) Raman spectra of pristine $WSe_2$ and V-doped $WSe_2$ with different concentrations (left panel). Raman mapping of the V-peak at 200 $cm^{-1}$ of 25% V-doped $WSe_2$ crystals with its line profile (right panel). (c) XPS of W 4*f*, Se 3*d*, and V 2*p* core-level electrons. (d) (left) ADF STEM image of microscale V-doped $WSe_2$ sample on the TEM grid. The inset shows a selected area electron diffraction pattern of the [001]-oriented sample, which confirms the hexagonal lattice structure of $WSe_2$. (middle) Atomic resolution ADF-STEM image of the V-doped $WSe_2$. (right) EDX spectra of V $K_\alpha$ (4.949 keV), W $M_\alpha$ (1.774 keV), and Se $L_\alpha$ (1.379 keV) edges acquired from the V-doped $WSe_2$ sample.

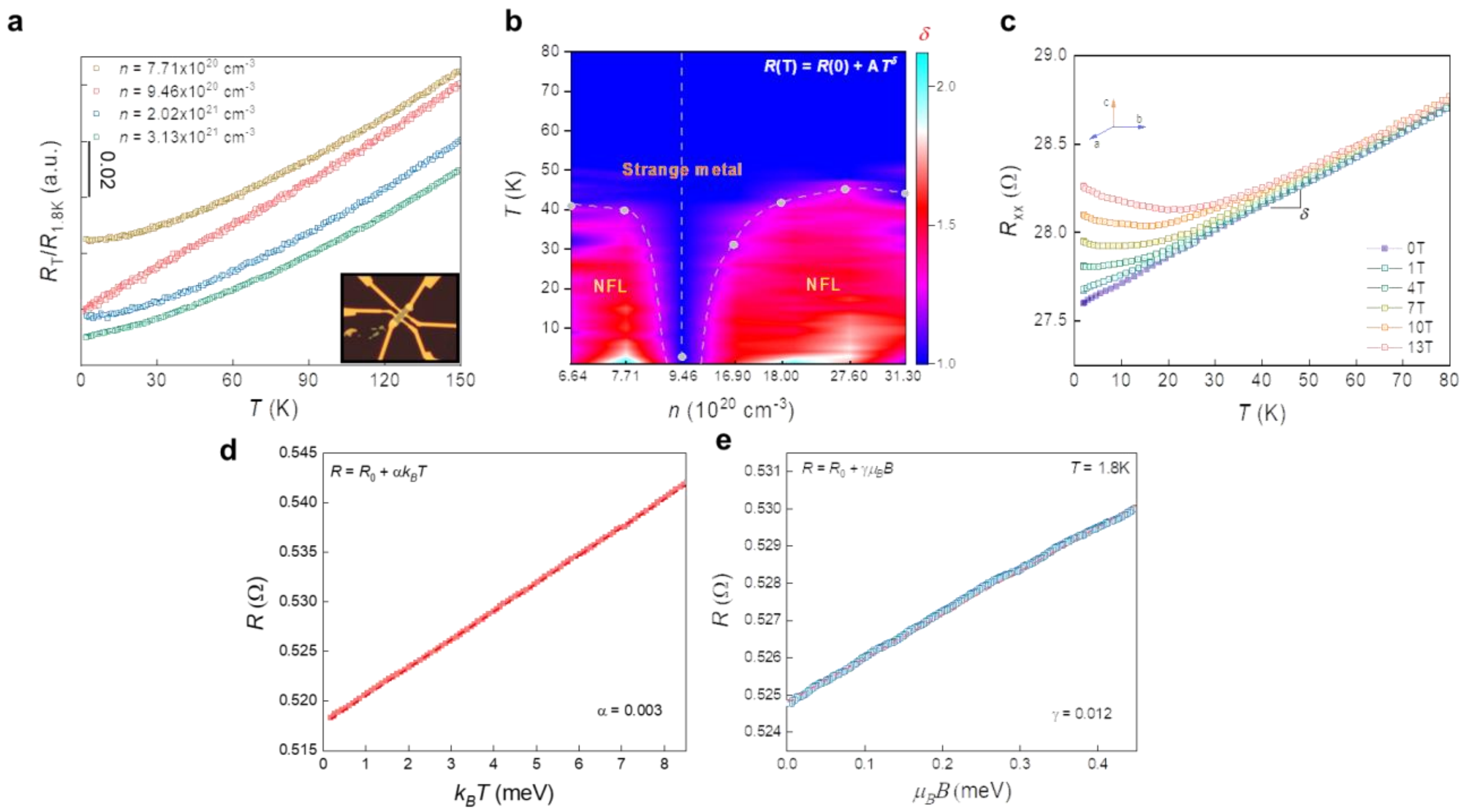

FIG. 4. *T*-linear and unsaturated *B*-linear resistance under a perpendicular magnetic field. (a) Temperature-dependent resistance of V-doped $WSe_2$ with different carrier concentrations. The strange metal is revealed in a critical carrier density of 9.46 x $10^{20}$ $cm^{-3}$. (b) Color map of $\delta$, the power factor of the $R(T) = R_0 + AT^{\delta}$, manifests the phase transition between strange metal and non-Fermi liquid with different concentrations. (c) Temperature dependence of resistance under magnetic field. (d-e) T-linear (d) and B-linear (e) resistances of another sample in the plot with energy scales $k_B$T and $\mu_B$B.

| Materials | | Type of band structure | Scaling factor $\gamma/a$ | Features | Ref. |
|---|---|---|---|---|---|
| **Cuprate** | $YBa_2Cu_3O_{7-\sigma}$ (YBCO) | d – orbital | 2 | - Strange metal<br>- Superconductivity<br>(Electron/Hole doped Mott insulator) | [12,26,45] |
| | $YBa_2Cu_4O_8$ (Y124) | | - | | [13,46] |
| | $La_{2-x}Ce_xCuO_4$ (LCCO) | | - | | [10,25,47] |
| | $La_{2-x}Sr_xCuO_4$ (LSCO) | | 1 | | [9,10,47] |
| | $Tl_2Ba_2CuO_{6+\delta}$ (Tl2201) | | 1 | | [27,48] |
| **Heavy Fermion** | β-$YbAlB_4$ / β-$YbAl_{1-x}Fe_xB_4$ | f – orbital | - | - Strange metal/Superconductivity (metallic) | [15] |
| | $YbNi_4(P_{1-x}As_x)_2$ | | | - Strange metal (Metallic ferromagnet) | [49] |
| | $CeRh_6Ge_4$ | | | - Strange metal by pressure (Ferromagnet) | [23,50] |
| **Manganites $La_{1-x}A_xMnO_3$** | | d -orbital | - | Strange metal (Metallic compound) | [51] |
| **Strontium Tinanate $SrTiO_{3-\delta}$** | | d -orbital | - | Strange metal (Metallic compound) | [52] |
| **Twist Bi-layer Graphene** | | d -orbital | - | Strange metal/Superconductivity (Correlated insulator) | [20–22] |
| **$BaFe_2(As_{1-x}P_x)_2$** | | d -orbital | 1 | Strange metal/Superconductivity (Antiferromagnetic metal) | [16,17,44] |
| **Ruthenate $Sr_3Ru_2O_7$** | | d -orbital | - | Strange metal by magnetic field (Highly 2D metallic) | [18,19,53] |
| **$V_xW_{1-x}Se_2$** | | d -orbital | 4 | Strange metal (Hole-doped valley spin insulator) | |

TAB. 1. Strange metal behaviour of various systems

# Supplementary Material

# Observation of strange metal in hole-doped valley-spin insulator

Tuan Dung Nguyen[1,2†], Baithi Mallesh[1,2†], Seon Je Kim[,2†], Houcine Bouzid[1], Byeongwook Cho[1,2], Xuan Phu Le[1,2], Tien Dat Ngo[3], Won Jong Yoo[3], Young-Min Kim[1,2*], Dinh Loc Duong[1,2*], Young Hee Lee[1,2,4*]

[1]Center for Integrated Nanostructure Physics (CINAP), Institute for Basic Science (IBS), Suwon 16419, Republic of Korea

[2]Department of Energy Science, Sungkyunkwan University, Suwon 16419, Republic of Korea

[3]SKKU Advanced Institute of Nano-Technology (SAINT), Sungkyunkwan University, Suwon 16419, Republic of Korea

[4]Department of Physics, Sungkyunkwan University, Suwon 16419, Republic of Korea

[†]These authors contributed equally to this work

[*] Correspondence to: youngmk@skku.edu, ddloc@skku.edu, leeyoung@skku.edu

# I. METHODS

## 1. Sample Preparation

A single crystal of V-doped $WSe_2$ was synthesized by a two-step chemical vapor transport (CVT) process using iodine as a transport agent. The first step is for the growth of the polycrystalline sample, and the second step allows the growth of the single crystal. Highly pure reactant powders of tungsten (99.999%, Sigma Aldrich), selenium (99.998%, Sigma Aldrich), and vanadium (99.99%, Sigma Aldrich) were mixed and ground uniformly in an atomic ratio of $V_xW_{1-x}Se_2$ by weight. The mixed powder was compressed into a pellet before loading into quartz ampules for sealing under a high vacuum and annealing at 1000 °C for 24 h. The polycrystalline powder was collected and resealed after adding iodine as a transporting agent. The temperatures of the cold and hot ends were 935 °C and 1050 °C, respectively, and were maintained for over 200 h to facilitate sufficient diffusion and stabilization for the growth of the single crystal. The pristine sample was processed similarly, except that the reactant powders used were of tungsten and selenium only.

## 2. Device Fabrication and Characterization

*Device Fabrication*: Few-layered pristine $WSe_2$ and V-doped $WSe_2$ flakes were exfoliated onto a $SiO_2$/Si substrate using a Scotch tape. The samples were then spin-coated using polymethyl methacrylate, and the substrate was heated on a hot plate at 160 °C for 3 min. Finally, the source (S) and drain (D) contacts (Cr/Au of 5/50 nm) and the hall bar designs were patterned by e-beam lithography before metal deposition.

*Electrical Characterization*: The temperature-dependent electrical resistance $R$(T) of the devices was examined using a physical property measurement system (PPMS, Quantum Design) in conjunction with a Bridge, Rotator and Tg-Mag (BRT) connecting resistivity module and Keithley 4200 system. Hall measurements were performed under high vacuum (~$10^{-7}$ Torr) with a Hall-probe station (Lake Shore Cryotronics system CRX-VF) and Keithley 4200 system.

*Structural and Chemical Characterization*: ADF STEM imaging was performed to acquire nanoscale morphologies and atomic scale lattice structures of the V-doped $WSe_2$ samples using a double aberration-corrected TEM apparatus (ARM200CF, JEOL Ltd.) operating at 80kV. The detector angle for the ADF-STEM images was set to ~53–180 mrad with a probe semi-angle of ~23 mrad. The statistical noise in the background of the STEM images was reduced by the Wiener filtering process, which was implemented in Digital Micrograph software with a commercial plug-in software (HREM Filter Pro, HREM Research Ltd.). The EDX analysis of the V-doped $WSe_2$ was performed using a dual-type silicon drift detector (SDD, JED-2300T, JEOL Ltd.) with the effective X-ray sensing area used was 100 $mm^2$.

### 3. Optical Measurements

X-ray photoemission spectroscopy (K-Alpha, THERMO FISHER) was performed to determine the elemental composition of V-doped $WSe_2$. Confocal Raman spectroscopy was conducted using a Nanobase system with an excitation wavelength of 532 nm under high vacuum ($10^{-6}$ Torr).

# II. SUPPLEMENTARY FIGURES

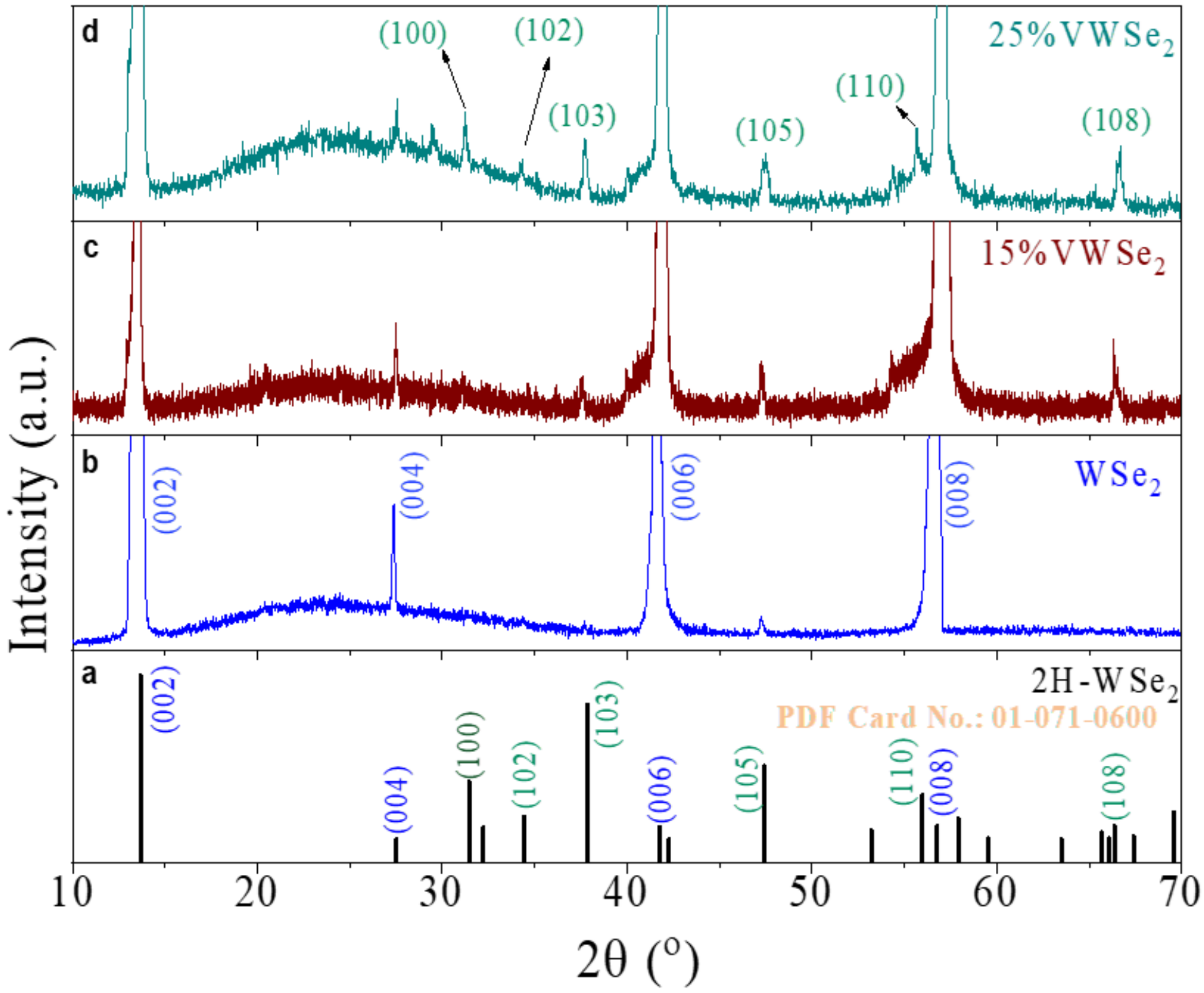

FIG. S1. Single crystal XRD spectra of pure $WSe_2$ and V-doped $WSe_2$. (a) Reference powder XRD patterns for 2H-$WSe_2$ adapted from ICSD database [1]. (b) Single crystal XRD pattern of the pure $WSe_2$. Several strong peaks at 13.6°, 41.7°, and 56.9° from the (002), (006), and (008) planes indicate a high crystallinity of 2H-$WSe_2$. (c) XRD pattern of 15% V-doped $WSe_2$. (d) XRD pattern of 25% V-doped $WSe_2$. The H phase of the host $WSe_2$ is displayed even with heavily V-doping concentration. The appearance of additional XRD peaks at 31.14° and 37.63° is also consistent with the (100) and (103) peaks of 2H-$WSe_2$. As the V doping concentration increases, the (002) peaks shifted to higher diffraction angles imply the shrink in lattice constant along the c axis, decreasing the interlayer distance.

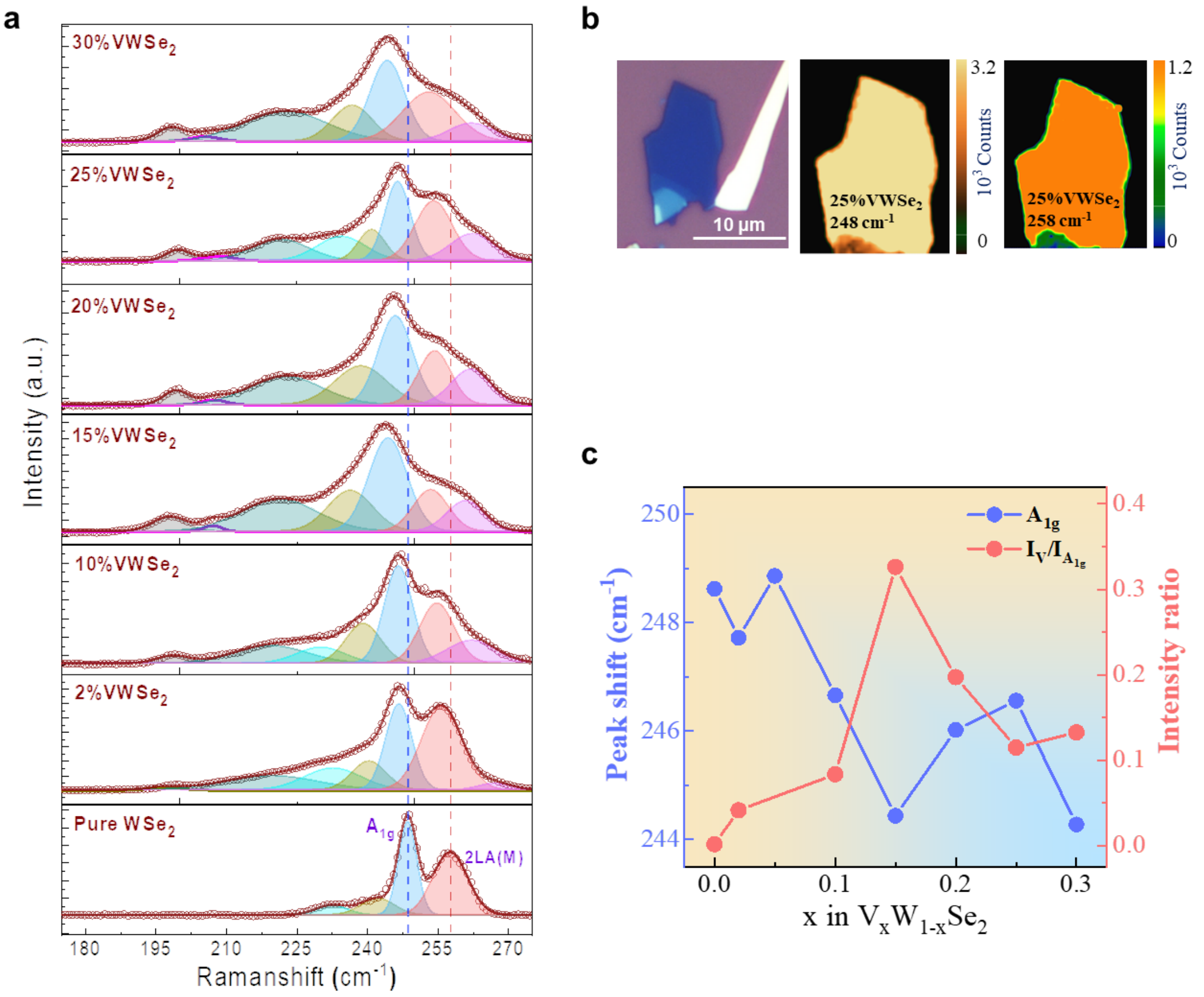

FIG. S2. Raman spectra of V-doped $WSe_2$. (a) Raman spectra of pure $WSe_2$ and V-doped $WSe_2$ with different V concentrations. (b) Optical image of exfoliated multilayer 25% V-$WSe_2$ and Raman mapping of the $A_{1g}$ and 2LA(M) modes peak at 248 $cm^{-1}$ and 258 $cm^{-1}$, respectively (top-right panel). (c) Blue shift of $A_{1g}$ and increase in V- mode dominance with an increase in x indicated by the intensity ratio of V-peak/$A_{1g}$ (bottom-right panel).

The Raman spectrum of pure $WSe_2$ reveals the in-plane $E_g$ and out-of-plane $A_{1g}$ peaks at 248 $cm^{-1}$ at an excitation wavelength of 532 nm. We observed an additional peak near 258 $cm^{-1}$ involving an LA phonon branch at point M (2LA(M)). Two more peaks featuring at 230 and 242 $cm^{-1}$ correspond to the phonon mode of $E_{1g}$-symmetry optical branch at K and M points of the Brillouin zone, respectively [2]. After doping, the Raman peaks are broadened, and more importantly, a distinct peak appears near 200 $cm^{-1}$, which becomes more prominent as the V-doping concentration increases from 2 to 30 %. This peak associated with V reflects in the V-Se bonding, indicating the incorporation of V atoms inside the $WSe_2$ host. To investigate the Raman spectrum in detail, the broadened peaks at V concentrations of 10% and 25% were fitted with more than 4 peaks, implying that the vibration modes appeared due to the presence of the V atoms. In FIG.S2 (b), we present the Raman mapping of 25% V-doped $WSe_2$. The appearance of peaks around 248 $cm^{-1}$ and 258 $cm^{-1}$ reveals the uniform distribution of Wand Se within the flake scale of the Raman spot size. Moreover, a blue shift of the $A_{1g}$ mode coupled with the enhancement in the intensity ratio of V and $A_{1g}$ modes of the V-peak at high V concentrations is observed in FIG. S2(c).

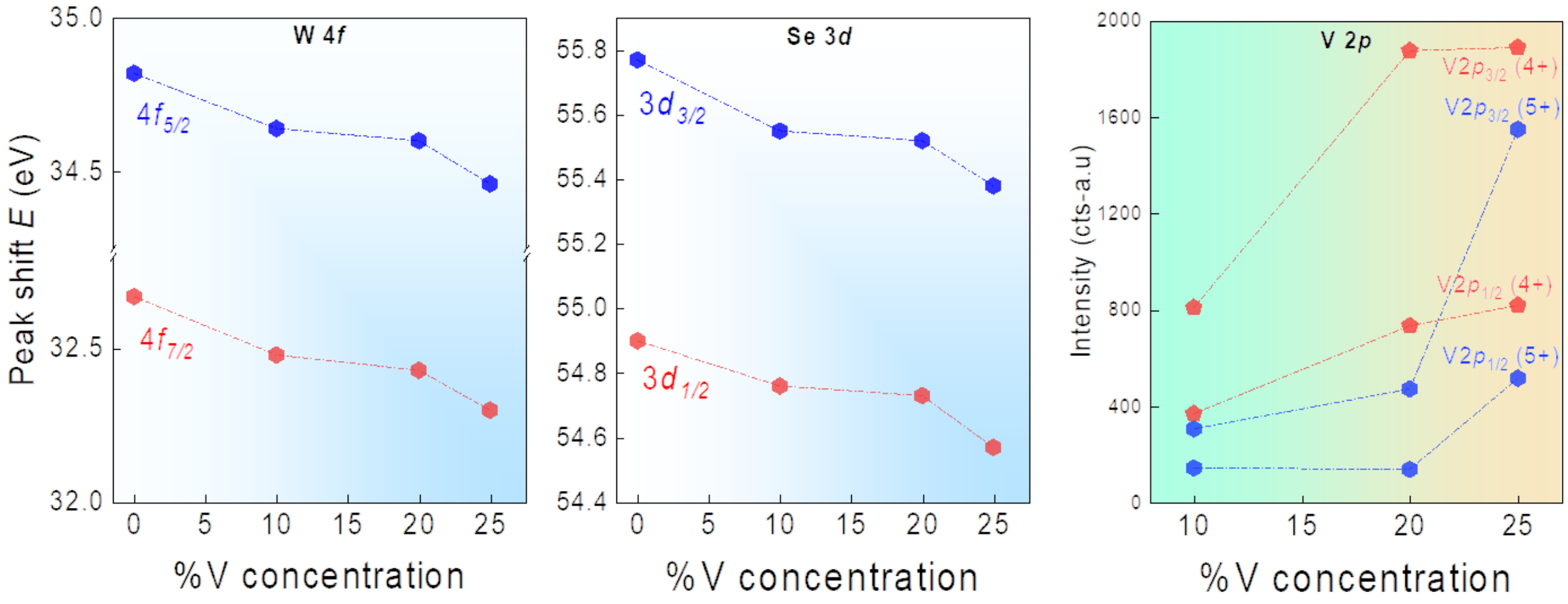

FIG. S3. Peak shift XPS of the W4*f*, Se3*d* and the intensity of V2*p* core-levels in the pure $WSe_2$ and V-doped $WSe_2$ samples. The doublets of W-4*f* (32.65 and 34.84 eV) and Se-3*d* (54.92 and 55.75 eV) core levels of pure 2H $WSe_2$ are red-shifted corresponding to V-doping concentration. The V-2*p* of vanadium atoms presents two pairs of V-2$p_{3/2}$ and V-2$p_{1/2}$ peaks at (513.7, 521.4 eV) and (516.4, 524.9 eV) in the V-doped $WSe_2$, respectively, and the intensity of the V-2*p* core level enhances as the V doping concentration increases, corresponding to V-Se bonding and indicating the successful doping of $WSe_2$ with V atoms. In addition, the position of W-*4f* and Se-*3p* is distinct from that of the 1T'-$WSe_2$ phase, confirming the H structure of the doped samples.

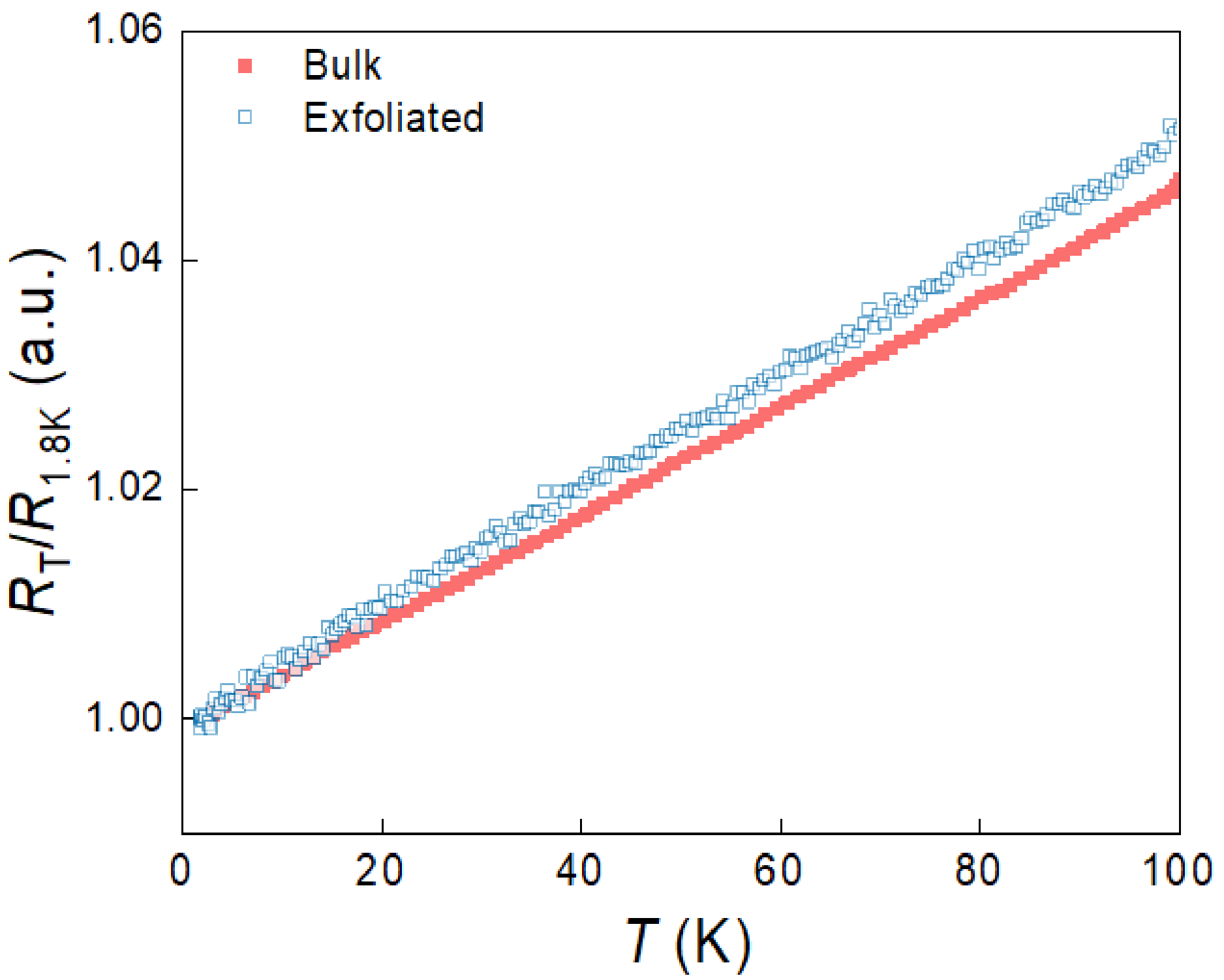

FIG. S4. Comparison of T-linear resistance at different samples. The blue curve from bulk samples with nominal 25% V-doped $WSe_2$. Two normalized curves are almost identical, indicating the same nature in both samples.